\documentclass[twocolumn,prl,amsmath,amssymb,superscriptaddress,showpacs]{revtex4-1}

\usepackage{graphicx}
\usepackage{dcolumn}
\usepackage{bm}

\def\reff#1{(\ref{#1})}

\begin{document}

\title{Extraordinary transmission induced by defects in semitransparent screens}

\author{V. Delgado}
\email{vdelgado@us.es}
\affiliation{Departamento de Electr\'onica y Electromagnetismo, Universidad de Sevilla, 41012-Sevilla, Spain}

\author{R. Marqu\'es}
\email{marques@us.es}
\affiliation{Departamento de Electr\'onica y Electromagnetismo, Universidad de Sevilla, 41012-Sevilla, Spain}

\author{L. Jelinek}
\email{l\textunderscore jelinek@us.es}
\affiliation{Department of Electromagnetic Field, Czech Technical University in Prague, 166 27-Prague, Czech Republic}

\date{\today}

\begin{abstract}
In this letter we present an analytical theory of Extraordinary optical transmission (EOT) through semi-transparent screens, such as thin metallic plates or high permittivity dielectric slabs. Using this theory we show that EOT appears not only for screens perforated by holes or slits, but also for screens loaded by any defects, including opaque defects. These results widen the scope of EOT concept, opening up the way to the study of new physical effects.
\end{abstract}


\maketitle


Extraordinary Optical Transmission (EOT) was first reported at optical frequencies by Ebbessen and co-workers \cite{Ebbesen-1998} in metallic screens perforated by a periodic array of sub-wavelength holes. Since then, EOT has been also reported at microwave frequencies \cite{Beruete-2004}, where metals behave as almost perfect conductors, and also in perforated lossy dielectric screens \cite{Sarrazin-2003,Lezec-2004}. EOT was also reported in 1D periodic structures, such as metallic screens perforated by a periodic array of slits \cite{Porto-1999}. First theories of EOT invoked the coupling between ``surface plasmons'' excited at both sides of the screen \cite{Ghaemi-1998,Pendry-2004}, in order to explain this effect, although other theories can be found in the literature, including generalization of Bethe's aperture theory \cite{Gordon-2007} and circuit models \cite{Medina-2008}. The reader may consult the excellent reviews by {C. Genet} et al. \cite{Genet-2007} or by {F. J. Garcia de Abajo} \cite{Abajo-2007} in order to have a complete overview of the topic.

\begin{figure}
\centering
\includegraphics[width=0.9\columnwidth]{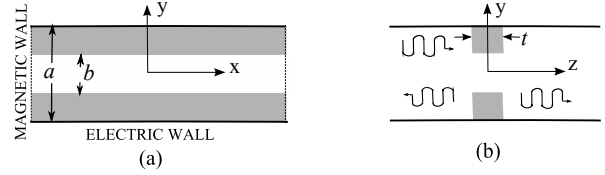}
\caption{\label{Fig1} Front and side views of the unit cell of the analyzed structure. Due to the polarization of the incident wave and to periodicity, upper and lower planes of the unit cell are virtual perfect conducting plates (PEC), so that the whole structure is equivalent to a symmetrical metallic iris in a parallel plate PEC waveguide.}
\end{figure}

Recently, the authors of this paper presented an analytical theory of EOT through perfect conducting screens \cite{Marques-2009} which was later extended to screens made of realistic conductors \cite{Delgado-2010}. This last generalization was based on the surface impedance concept which is widely used in classical electromagnetism in order to analyze skin effect on imperfect conductors \cite{Jackson01}. For thin metallic screens it is more convenient to use the generalization of this concept, providing a matrix relation between electric and magnetic fields at both sides (``$^+$'' and ``$^-$'') of the screen \cite{Tretyakov} . This matrix can be diagonalized in the form \cite{Delgado-2010}
\begin{eqnarray}\label{f1}
\left[ {\begin{array}{*{20}c}
   {{\mathbf{E}}_\parallel ^ +   + {\mathbf{E}}_\parallel ^ -  }  \\
   {{\mathbf{E}}_\parallel ^ +   - {\mathbf{E}}_\parallel ^ -  }  \\

 \end{array} } \right]\!\! \approx\!\! \left[ {\begin{array}{*{20}c}
   {Z^{\left( 1 \right)} } & 0  \\
   0 & {Z^{\left( 2 \right)} }  \\

 \end{array} } \right]\!\!\left[ {\begin{array}{*{20}c}
   {\left( {{\mathbf{H}}_\parallel ^ +   - {\mathbf{H}}_\parallel ^ -  } \right) \times {\mathbf{n}}}  \\
   {\left( {{\mathbf{H}}_\parallel ^ +   + {\mathbf{H}}_\parallel ^ -  } \right) \times {\mathbf{n}}}  \\

 \end{array} } \right]\!\!,
\end{eqnarray}
where ${{\mathbf{E}}_\parallel ^{\pm} }$ and ${{\mathbf{H}}_\parallel ^{\pm} }$ are the electric and magnetic field components parallel to the screen, ${\bf n}$ is the unitary vector normal to the screen, and ${Z^{\left( 1 \right)} }$, ${Z^{\left( 2 \right)} }$ are some surface impedances which depend on the characteristics of the screen. For plane waves impinging on an infinite and homogeneous screen this expression can be made exact (then ${Z^{\left( 1 \right)} }$ and ${Z^{\left( 2 \right)} }$ depend on the angle of incidence and polarization), as well as for slab discontinuities in a hollow metallic waveguide, as that shown in Fig.\ref{Fig1} (in this case ${Z^{\left( 1 \right)} }$ and ${Z^{\left( 2 \right)} }$ are different for each mode). For screens perforated by small holes or loaded by other small defects, \reff{f1} can be still considered as an useful approximation \cite{Delgado-2010}. Finally, \reff{f1} can be still formally used for relating fields at both sides of a channel in the screen. In this case ${Z^{\left( 1 \right)} }$ and ${Z^{\left( 2 \right)} }$ are functions of the impedance of the dominant mode in the channel \cite{Delgado-2010}.

\begin{figure}
\centering
\includegraphics[width=0.9\columnwidth]{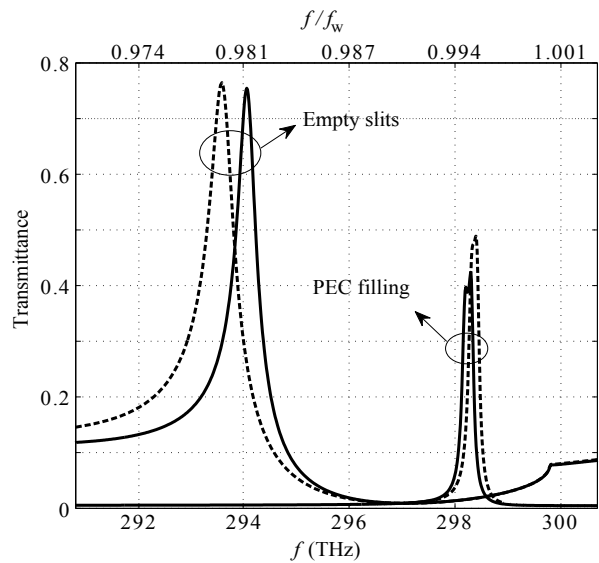}
\caption{\label{Fig2} Transmittance through an Ag screen of thickness $t=100$ nm, periodically perforated by an array of parallel slits of periodicity $a=1\;\mu$m and width $b=a/6$, which can be either empty or filled by a PEC. For the permittivity of Ag screen we used a Drude model with the same constants already used in \cite{Kady-2000} and the frequency of collisions corrected according to Eq.2.30 of \cite{Coronado-2003}, in order to take into account collisions of electrons with the boundaries of the screen. For empty slits we assumed an extra width equal to two times the penetration depth in the Ag boundaries. Solid lines are our results computed from \reff{e1} - \reff{e4}. Dashed lines are results from CST. Upper scale shows the ratio $f/f_{\rm w}$, where $f_{\rm w}$ is the frequency corresponding to Wood's anomaly $f_{\rm w}=c/a$.}
\end{figure}

In order to simplify the analysis we will first consider EOT through an array of parallel slits perforated in a metallic screen \cite{Porto-1999}. The unit cell of this structure is shown in Fig.\ref{Fig1} and, as usual, normal incidence of an electromagnetic wave with polarization perpendicular to the slits ($E_{\rm y}$) is considered. The particularization of the theoretical model to this structure has been reported elsewhere \cite{Delgado-2010a,Delgado-2010b} and will not be repeated. Here, for the sake completeness, we will only write the final equations:
\begin{multline}
(1+Z_{n}^{(1)} Y_n)(T_n+R_n) = \\ 2(1+R+T) - 2Z_{0}^{(1)} Y_0(1-R-T) \;,\;\; n=1,2... N \label{e1}
\end{multline}
\begin{multline}
(1+Z_{n}^{(2)} Y_n)(T_n-R_n) = \\ 2(-1-R+T) - 2 Z_{0}^{(2)} Y_0(-1+R-T) \;,\;\; n=1,2... N \label{e2}
\end{multline}
\begin{multline}
(1+R+T+Z_{s}^{(1)} Y_0T-Z_{s}^{(1)} Y_0+Z_{s}^{(1)} Y_0 R) + \\
\sum_{n=1}^N (T_n+R_n+Z_{s}^{(1)} Y_n T_n+Z_{s}^{(1)} Y_n R_n) \text{sinc}\left(\frac{bn\pi}{a}\right) = 0 \label{e3}
\end{multline}
\begin{multline}
(-1-R+T+Z_{s}^{(2)} Y_0 T + Z_{s}^{(2)} Y_0 - Z_{s}^{(2)} Y_0 R) + \\
\sum_{n=1}^N (T_n-R_n+Z_{s}^{(2)} Y_n T_n -Z_{s}^{(2)} Y_n R_n) \text{sinc}\left(\frac{bn\pi}{a}\right) = 0. \label{e4}
\end{multline}
These equations are valid for small slits ($b\lesssim a/4$), and must be solved for the transmission and reflection coefficients $T$ and $R$, and for the auxiliary coefficients $T_n$ and $R_n$, with $N\sim a/b$. In \reff{e1}-\reff{e4}, $Y_n$ are the admittances of the even TM$_{2n}$ modes of the empty parallel plate waveguide of Fig.\ref{Fig1}, $Z_{n}^{(1)}$ and $Z_{n}^{(2)}$ are the surface impedances \reff{f1} of the screen for each mode, and $Z_{s}^{(1)}$, $Z_{s}^{(2)}$ are the surface impedances \reff{f1} associated with the fundamental TEM mode of the slit. It is worth noting that $N$ in \reff{e1}-\reff{e4} is usually not more than $8$ for typical widths of the slits, so that they provide a fast and accurate analytical tool for computation. From another point of view, Eqs. \reff{e1} to \reff{e4} show that, for a given screen and periodicity, transmittance is determined by the impedances $Z_{s}^{(1)}$ and $Z_{s}^{(2)}$. Therefore, a natural question is how the transmittance changes with these impedances. In electromagnetism there are two limits for surface impedance, one of them $Z\rightarrow 0$ corresponding to perfect electric conductor (PEC) and the other $Z\rightarrow\infty$ corresponding to perfect ``magnetic'' conductor (PMC). Both of them correspond to opaque surfaces, but, for thin semi-transparent screens it is worth to ask what will be the effect of filling the slit with PEC or PMC. Fig.\ref{Fig2} shows the transmittance through a thin screen of Ag, periodically perforated by parallel slits which are either empty or filled by PEC. The results computed from \reff{e1}-\reff{e4} (solid lines) are compared with the results computed using the commercial electromagnetic solver CST Microwave Studio (dashed lines) showing very good agreement. In both cases an extraordinary transmission peak appears close to the Wood's anomaly. Computations (not shown) made for slits filled with PMC provide similar results. These results strongly suggest that EOT is actually a much more general effect than high transmittance through periodic sub-wavelength channels, also appearing for opaque defects in semi-transparent screens.

\begin{figure}
\centering
\includegraphics[width=0.9\columnwidth]{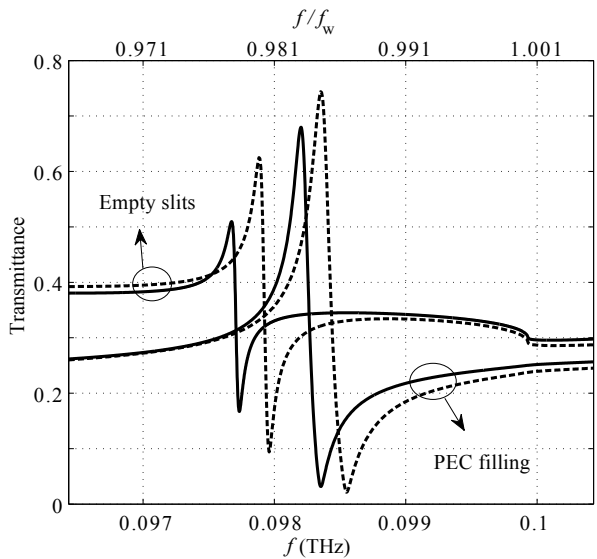}
\caption{\label{Fig3} Transmittance through a zirconium-tin-titanate ($\varepsilon = 92.7(1 + 0.005i)$) screen of thickness $t=0.12$ mm, periodically perforated by an array of parallel slits of periodicity $a=3$ mm and width $b=a/6$, which can be either empty or filled by a PEC. Solid lines are our results computed from \reff{e1} - \reff{e4}. Dashed lines are results from CST. Upper scale shows the ratio $f/f_{\rm w}$, where $f_{\rm w}$ is the frequency corresponding to Wood's anomaly $f_{\rm w}=c/a$.}
\end{figure}

Neither PEC nor PMC are available, even approximately, at optical frequencies. Therefore, the previous results are just theoretical calculations. In order to show a more realistic example, we will analyze transmittance through a high permittivity dielectric slab. In this case, except at Fabry-Perot resonances, a very low transmittance is expected due to the strong impedance discontinuity at the boundaries of the screen. However, if the screen is periodically perforated by an array of periodic parallel slits, an extraordinary transmission peak appears near Wood's anomaly, just as in EOT through metallic screens. This effect is shown in Fig.\ref{Fig3} which shows transmittance through a periodic array of slits perforated in a slab of zirconium-tin-titanate with $\varepsilon = 92.7(1 + 0.005i)$ \cite{Bolivar-2003}. As in the previous example, transmittance through the same array of slits filled with a PEC (most metals behave as PEC at such frequencies) is also shown in the Figure. Besides the good agreement between our analytic results and electromagnetic simulations, the figure also shows how EOT can be induced by periodic opaque defects in semi-transparent screens made of high permittivity dielectrics.

\begin{figure}
\centering
\includegraphics[width=0.9\columnwidth]{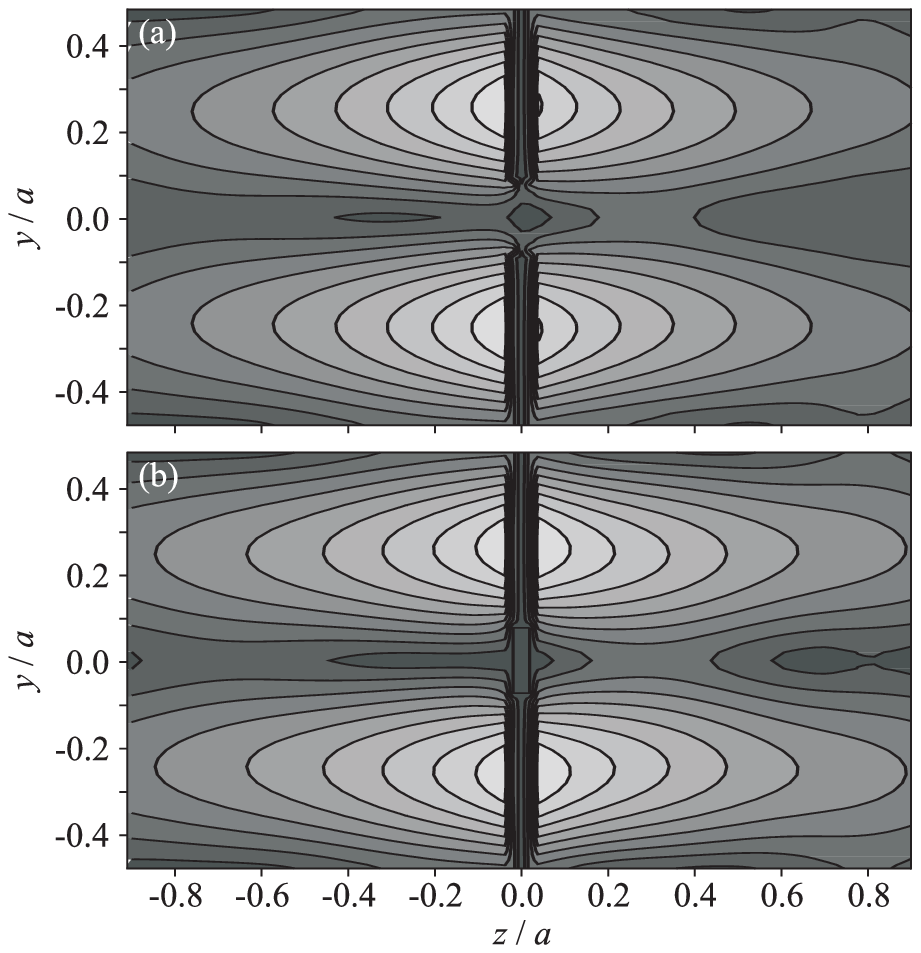}
\caption{\label{Fig4} Normalized electric field distribution (absolute value) at both sides of the screen for the configuration analyzed in Fig.\ref{Fig3} with empty slits (a) and slits filled by PEC (b). Calculations were made using CST, and correspond to the frequency of maximum transmission in both cases. Black color corresponds to zero field and white color to maximum field value.}
\end{figure}

According to standard theories about EOT, this effect is associated to the excitation of weakly coupled leaky surface waves at both sides of the screen, with transverse wavenumbers close to $2\pi/a$, i.e., with wavelengths close to the periodicity. In standard EOT, these surface waves are coupled through the slits (or the holes in 2D systems) perforated in the screen. For semi-transparent screens, this coupling can be made through the screen itself and the role of the defects is merely to allow for the excitation of the surface waves. In order to support this interpretation, we show in Fig.\ref{Fig4} the electric field distribution at both sides of the screen in the configuration of Fig.\ref{Fig3} at the frequency of maximum transmission. A clear standing wave pattern corresponding to the simultaneous excitation of two surface waves traveling at opposite directions along each side of the screen can be appreciated in both cases.

\begin{figure}
\centering
\includegraphics[width=0.9\columnwidth]{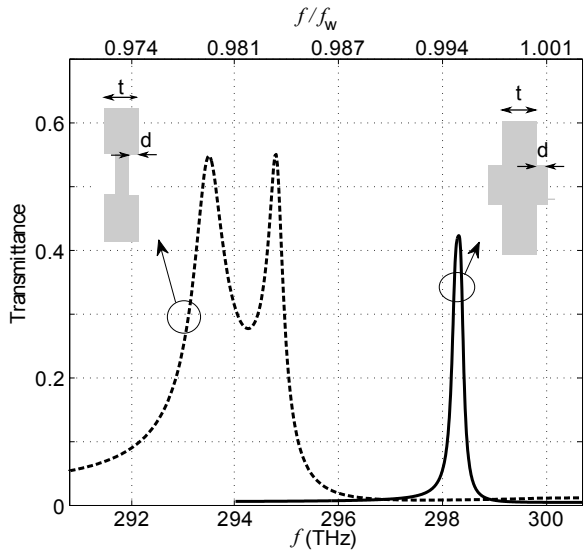}
\caption{\label{Fig5} Transmittance through an Ag screen of thickness $t=100$ nm, periodically loaded by an array of grooves/ridges of periodicity $a=1\;\mu$ m, width $b=a/6$, and depth/height $d=25$ nm (see the insets). Transmittance curves have been computed using CST Microwave Studio.}
\end{figure}

Returning to optical frequencies, we will now investigate how EOT can be induced by other defects than PECs or PMCs. Specifically, we will consider EOT induced by ridges or grooves in Ag screens. In this case, the analytical theory summarized in \reff{e1}-\reff{e4} can not be applied. However, the system can be still solved by using a commercial electromagnetic solver. In Fig.\ref{Fig5} EOT through a silver screen similar to that analyzed in Fig.\ref{Fig2}, but with ridges or grooves instead of slits is analyzed using CST Microwave Studio. In both cases sharp transmission peaks appear near Wood's anomaly, thus confirming the hypothesis. The main difference between these results and those shown in previous figures is the presence of two peaks for the groove configuration (in fact, both peaks also appear in the ridge configuration for thinner screens). The presence of two extraordinary transmission peaks in thick metallic screens is not new \cite{Abajo-2007} and, for lossless PEC screens, has been explained in the frame of circuit theories \cite{Medina-2008} and analytical models \cite{Marques-2009}. The electric field distribution (not shown) at both sides of the screen for the three transmission peaks of Fig.\ref{Fig5} (ridge and groove configuration) exhibits very similar behavior to that of Fig.\ref{Fig4}. Moreover, whenever two peaks appear, the relative phase of fields at both sides of the screen is different for each peak, showing a symmetric or antisymmetric pattern, which correspond to the even and odd couplings already discussed in \cite{Medina-2008} among others. These observation confirms that in all cases the physical mechanism behind transmission is very similar to standard EOT mechanism.

\begin{figure}
\centering
\includegraphics[width=0.9\columnwidth]{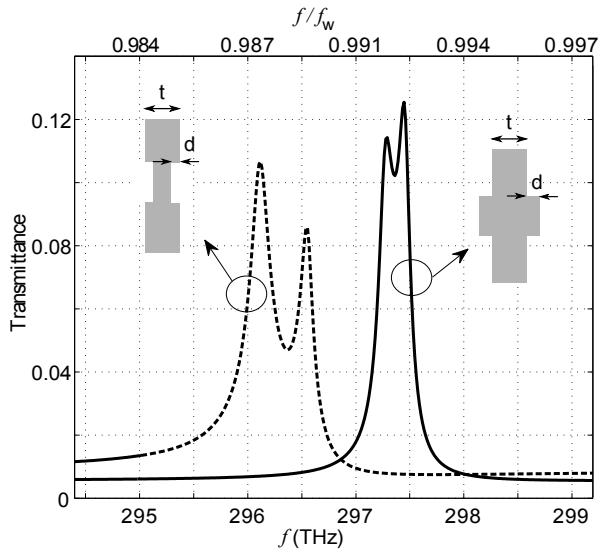}
\caption{\label{Fig6} Transmittance through an Ag screen of thickness $t=100$ nm, periodically loaded by an square array of square protuberances of periodicity $a=1\;\mu$m, width $b=a/4$, and depth/height $d=25$nm (see the inset). Transmittance curves have been computed using CST Microwave Studio.}
\end{figure}

Until now we have considered 1D configurations. Next, we will investigate if a similar effect can be observed in 2D configurations. For this purpose we will use again a commercial electromagnetic solver in order to compute transmittance through silver screens loaded by an array of 2D periodic protuberances. Square periodicity and square protuberances have been considered in order to simplify the analysis. Fig.\ref{Fig6} shows the transmittance through a semi-transparent silver screen loaded with protuberances. The analysis of field distribution (not shown) provide results very similar to those of Figs.\ref{Fig4}.

From previous analysis and results we can conclude that the same physical mechanism behind EOT through opaque screens - that is, the excitation of weakly coupled surface waves at both sides of the screen with wavelengths close to the periodicity - is also behind extraordinary transmission peaks in 1D and 2D semi-transparent screens loaded by periodic defects. This fact has been shown by means of an analytical theory which takes into account all these effects, and by careful numerical simulations. Analytical theory shows that all these effects are predicted by the same set of equations \reff{e1}-\reff{e4}, by simply including in the model the appropriate surface impedance for the slits and/or the defects. Numerical simulations have shown not only the accuracy of the analytical theory, but also that the field configurations at both sides of the screen are very similar in all cases. We feel that the reported results will help to a better understanding of EOT, extending the physical concept of extraordinary transmission to new and interesting physical phenomena.

\begin{acknowledgments}
This work has been supported by the Spanish Ministerio de Educaci\'on y Ciencia and European Union FEDER funds (project No. CSD2008-00066), by the Czech Grant Agency (project No. 102/09/0314), and by the Czech Technical University in Prague (project No. SGS10/271/OHK3/3T/13).
\end{acknowledgments}

\end{document}